# Destructive Read by Wave Interference for Arbitration

Rex Y. Li

Fortelee2016@gmail.com

*Abstract-* **With the advent of big data and deep learning, computation power has become a bottleneck for many applications. Network-on-Chip (NoC) has been proposed to enable multiprocessor acceleration for deep learning computation, and efficient arbitration is a key issue for high performance. In this paper, an arbitration scheme based on interference of wave is proposed. In this scheme, home node sends out multiple tokens in different frequencies, and nodes that are competing for bus can capture token by sending out wave that cancels its specific wave token. In this scheme, speed-of-light arbitration can be achieved, and full information is available to all nodes in arbitration.**

## I. INTRODUCTION

DEEP-Learning-based artificial intelligence (AI) has been popular for several years [1-5]. Currently, AI applications can help people with face detection, human-machine interaction [2], image processing [3] etc.

Computation power has become a bottleneck as deep learning has been proposed. In 2012, AlexNet [1] uses GPU to accelerate deep learning applications. Since then, using many processors to enhance algorithm deployment performance has been widely accepted. Network-on-Chip [6] is a method to integrated large-scale processor network to accelerate computation. In NoC, arbitration is a key problem to decide which processor node can obtain resource. Conventional arbitration scheme needs large bandwidth and induces low latency, and can degrade performance of NoC for AI applications. In this paper, a new arbitration scheme is proposed based on interference of electromagnetic waves. Unlike conventional arbitration where only basedband frequency band is used, this arbitration scheme takes advantage of full frequency spectral resource. In addition, since wave interference takes place at speed-of-light, this arbitration scheme can achieve ultra high speed arbitration.

## II. MODULATION AND INTERFERENCE

All electronic signals are in essence electromagnetic (EM) waves. Every EM wave can be expressed as [7]:

$$S(t)=A(t)\cos(\omega(t)t+\theta(t))$$

where $A(t)$ is variable amplitude for EM wave, $\omega(t)$ is frequency and $\theta(t)$ is phase.

We can mount information on EM wave by modulation. By modulation, we can change the amplitude (A), frequency ($\omega$) and/or phase of EM wave. For example, when we want to transfer bit '1', we can set A(t) to 1 V, while when we want to transfer bit '0', we can set A(t) to 0. This is amplitude modulation, equivalent to on-off keying (OOK). In order to make EM wave transfer signal properly, the frequency components in information (A(t)) cannot have too high frequency ($<<\omega$).

In order to retrieve information from modulated EM wave, we can use demodulation. Demodulation for amplitude modulation is simple and straightforward: we just multiply the received EM wave with the EM wave with same frequency and phase, but with unit amplitude. After integration, the output is:

$$\int_0^T A\cos(\omega t + \theta) \cdot \frac{2}{T}\cos(\omega t + \theta)dt = A$$

Therefore amplitude A (with information) is retrieved.

Another useful characteristic of modulation and demodulation is that EM waves with different frequencies are orthogonal to each other. For example, two carriers with frequency $\omega_1$ and $\omega_2$ is orthogonal since when we use $\omega_2$ to demodulate $\omega_2$, we get zero:

$$\int_0^T A\cos(\omega_1 t + \theta) \cdot \frac{2}{T}\cos(\omega_2 t + \theta)dt = 0$$

Another important characteristic of EM wave is interference. Two waves with same frequency but different phase can enhance or cancel each other:

$A_1(t)\cos(\omega t+\theta_1)+ A_2(t)\cos(\omega t+\theta_2) = [A_1(t)+ A_2(t)]\cos(\omega t+\theta_1)$

if $\theta_1=\theta_2$;

$A_1(t)\cos(\omega t+\theta_1)+ A_2(t)\cos(\omega t+\theta_2) = [A_1(t) - A_2(t)]\cos(\omega t+\theta_1)$

if $\theta_1=\theta_2+\pi$.

Interference is a common phenomenon for all waves, and can be observed in daily life (Fig. 1). It also exists for EM waves, and can be used for signal manipulation.

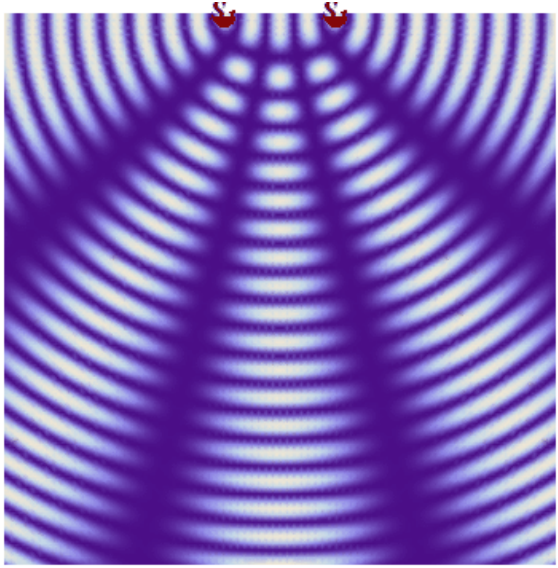

Fig. 1. Interference of waves.

### III. PROPOSED SCHEME

Arbitration is one of the most important operations in NoC. In [6], a photonic-based system is used for arbitration. Token is released by home node, and when one node captures the token, it would eliminate that token by resonator. By doing this, destructive read is achieved.

However, silicon photonic is still expensive and its yield is low. In this section, a practical destructive read scheme with wave interference is proposed. The system diagram is shown in Fig. 2.

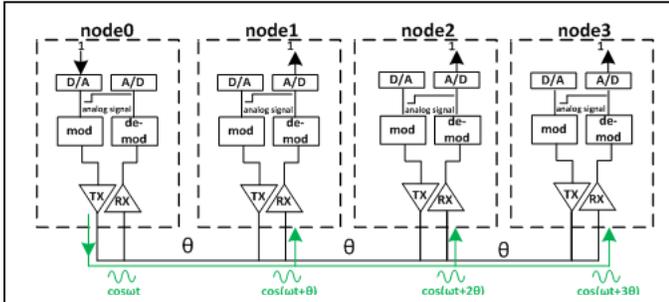

Fig. 2 System diagram

In the proposed system, each node is coupled to the same transmission line (TL). TL can be implemented with regular metal wire in traditional CMOS process, and is low cost. For example, home node (node 0) sends out token at frequency of $\omega$, and this token EM wave can be expressed as $A\cos\omega t$. Node 1, node 2 and node 3 are placed with equal electrical distance. Therefore, the token EM wave reaches node 1 with phase of $\theta$, at node 2 with phase of $2\theta$, and reaches node 3 with phase of $2\theta$.

In order to capture token, node 1 can release a EM wave with amplitude of A, but phase of $(\theta+\pi)$ after it detects token. This signal will propagate to node 2 and node 3 with phase of $(2\theta+\pi)$, $(3\theta+\pi)$.

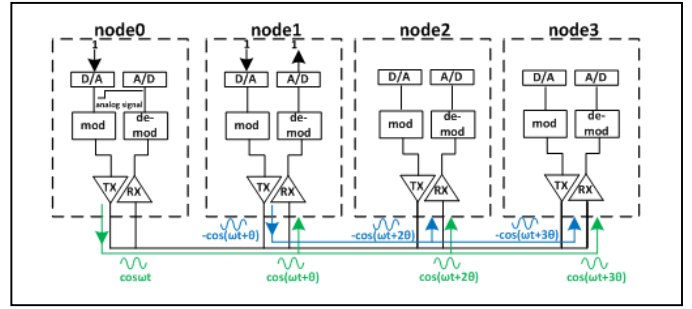

Fig. 3 Cancellation propagation of token.

Since for EM wave, we have:

$$\cos(\omega t+\theta) + \cos(\omega t+\theta+\pi) = 0$$

Therefore at node 2 and node 3, token EM wave and EM wave released by node 1 (capturing wave) cancels each other. Node 2 and node 3 cannot detect token. Equivalently, token is destructively read by node 1.

By doing this, a basic daisy chain arbitration scheme is achieved. Node 1 has the highest priority while node 3 has the lowest priority. When node i wants to use bus, it releases capturing wave when token arrives, and it keeps silent if it does not want to use bus.

Basic daisy chain is easy to implement, however it is takes time to finish arbitration and it lacks flexibility. Nodes cannot release token simultaneously, which induces latency. In order to improve performance, parallel daisy chain can be implemented.

In parallel daisy chain arbitration, for k nodes (n1, n2,…nk) in arbitration, split one token into k tokens (t1, t2,…tk) with orthogonal carriers $\omega 1$, $\omega 2$,… $\omega k$. Node n0 (home node) broadcasts all tokens t1~tn at the same time. n1 has the highest priority, nk has the lowest priority. Every node can read all tokens by proper demodulation. In addition, even more flexibility is possible if home node sends more tokens, say, 2n tokens.

Node ni can only cancel token ti with token cancellation signal with carrier frequency $\omega i$. Node ni needs tokens t1, t2,…ti to win arbitration. It wins when no nodes with higher priority competes with it. Virtually, k arbitrations run in parallel within one interconnect, and one arbitration runs at each orthogonal carrier channel. There is no interference between different tokens since multiple arbitrations are in different carrier channels. Therefore token cancellation signals can be sent simultaneously, and speed-of-light arbitration can be achieved.

In addition to fast arbitration, information is complete in this scheme. Competing nodes know who wins arbitration at once based on whether they get token(s). Upstream node also knows who takes the token based on received backward traveling wave phase/magnitude.

As an example, we still have three nodes. Three nodes n1, n2, n3 compete for bus, and we have priority of n1> n2> n3. Home node n0 broadcasts tokens t1, t2, t3 at carrier frequencies ω1, ω2 and ω3. Node n1 can cancel token t1, node n2 can cancel token t2, t1, node n3 can cancel token t3. n1 needs t1 to win, n2 needs t1 and t2 to win, n3 needs all tokens t1, t2 and t3 to win. Suppose all nodes are interested in bus, then token t1 is received and blocked by n1. Node1 knows that it wins arbitration since it detected t1. Node2 and node3 know they lose since they did not detect t1 (Fig. 4).

on the amplitude and phase of received signal in three carriers. Therefore, information is complete for all nodes. This is different from arbitration in [6], where only winner and home node knows the arbitration results. With complete information, each node (competing nodes and home node) can have statistics of arbitration, and can adjust priority properly. In addition, competing nodes can adjust their strategies accordingly to win next arbitration.

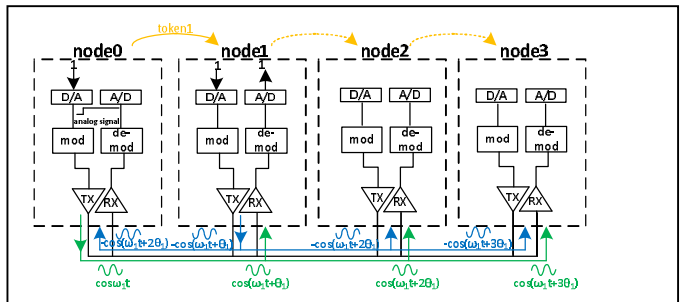

Fig. 4 Token t1 of parallel daisy chain arbitration.

For token t2, it is received and blocked by n2. However, n2 knows that it loses as it did not detect t1. n3 does not detect t2, it knows that n2 is also competing for bus. Also, backward traveling wave of cancellation signal by n2 goes to n1. Based on received signal magnitude and phase at carrier ω2, it knows that n2 is also competing for bus.

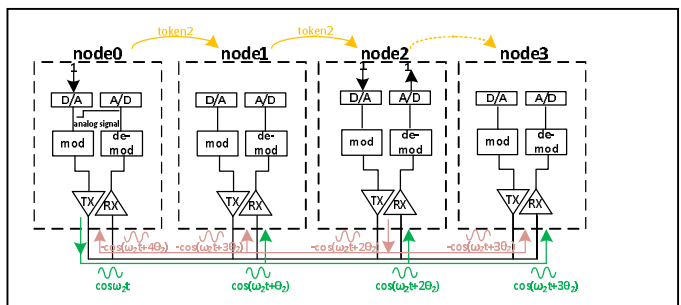

Fig. 5 Token t2 of parallel daisy chain arbitration.

As for token t3, it is received by n3. Backward traveling wave of cancellation signal by n3 goes to n1 and n2. Based on received signal magnitude and phase at carrier ω3, they know that n3 is also competing for bus.

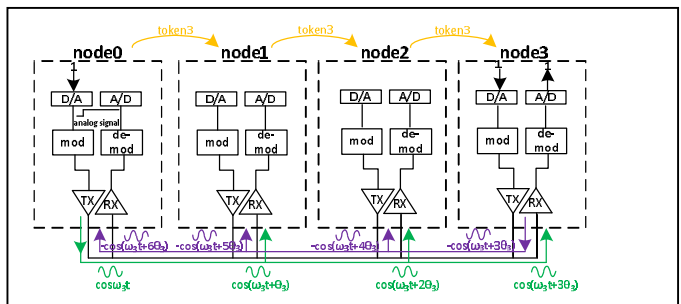

Fig. 6 Token t3 of parallel daisy chain arbitration.

In addition, backward traveling wave of cancellation signal by n1, n2 and n3 all go back to home node n0. Therefore, home node n0 knows that all three nodes are competing for bus based

## IV. SIMULATION RESULTS

The simulation is based on example that is given in previous section (three nodes competing for bus). Circuit-level simulation is done with Cadence Spectre as simulator. All nodes are capacitive-coupled to transmission line, and we used transmission line model in Cadence analogLib with imperfect termination to model non-ideality in real scenario.

We let node0 (home node) broadcast tokens at carrier frequencies t1:1 GHz, t2:2 GHz and t3:1.5 GHz. Nodes 1~3 are competing for bus. All are trying to get bus, so three token cancellation signals will be sent by node1, node2 and node3.

Node1 sends token cancellation signal of token1 (at 1 GHz). Therefore node2 and node3 will not sense token1 after it is cancelled, as shown in simulation results (Fig. 7 and 8). At node2 and node3, RF token signal at 1 GHz will be quenched after token cancellation. At node2 and node3, demodulated (analog) signal will become 0 mV after token cancellation.

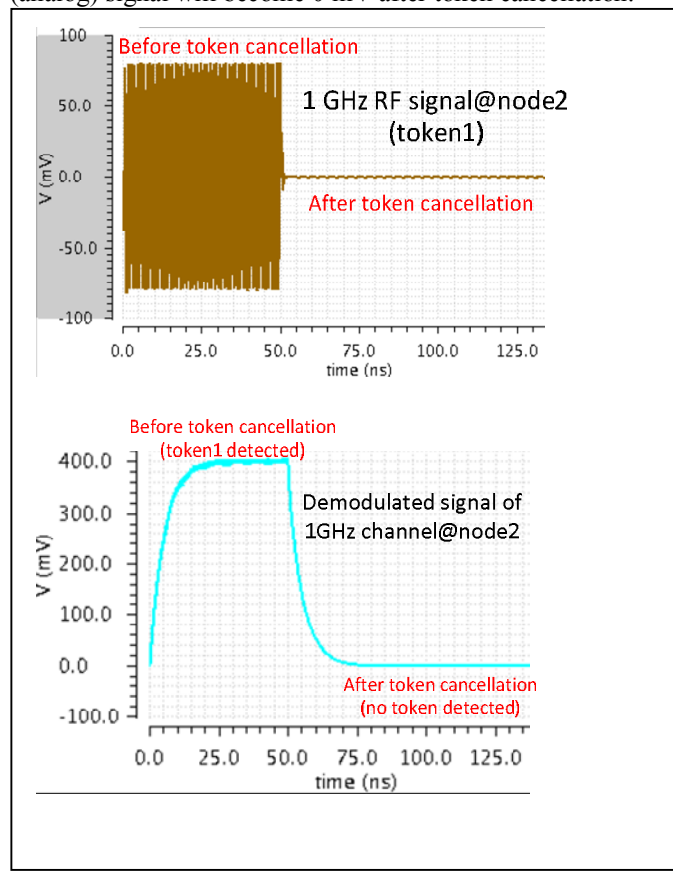

Fig.7 RF token 1 signal at node 2.

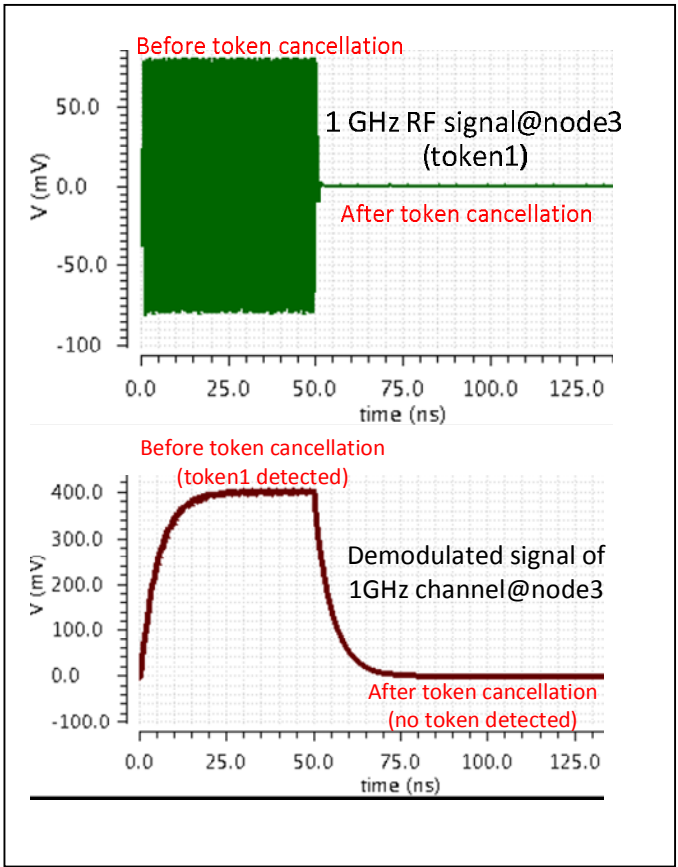

Fig. 8. RF token 1 signal at node 3.

For node 1, it can successfully detect and capture token 1 (Fig. 9). Similarly, node 2 and node 3 can successfully capture token 2 and token 3, respectively.

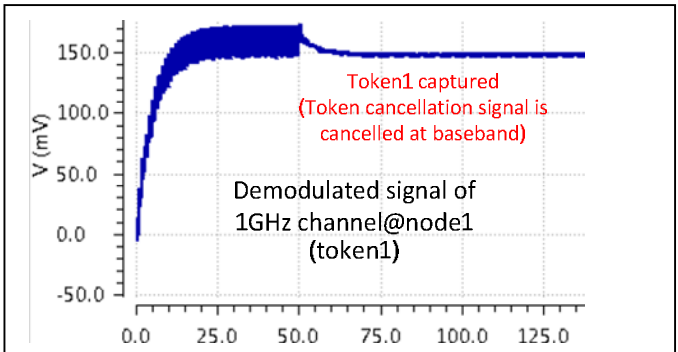

Fig. 9 Token t1 captured by node 1.

In addition, backward wave is also important. The token cancellation signal will travel backwards to upstream nodes. Fig. 10 shows backward traveling wave of 1GHz token cancellation signal, and the same result applies for other token cancellation signals at different carriers. Upstream nodes (including home node) know which downstream node is interested in bus, and strategy optimization would be possible with this information.

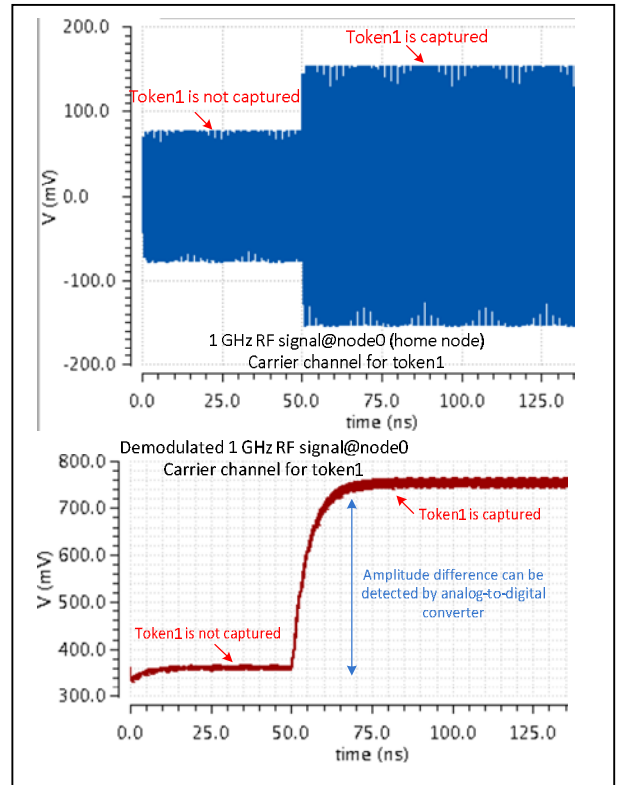

Fig. 10 Token t1 cancellation wave propagates backward to home node and is detected.

## V. CONCLUSION

In this paper, we proposed an arbitration scheme where information is fully available for all competing nodes. With the proposed destructive-read token (physically its token is 'wave'), all nodes know who wins arbitration based on which token is taken. Further, every node in arbitration knows its competitors. In game theory, every node has perfect information, thus optimal strategy is possible, and this provides more information to every node in arbitration game and they can have more efficient strategy based on its competitors. In turn, more efficient priority allocation is possible, and priority can be allocated based on current/history status of arbitration. Simulation verifies our implementation of EM arbitration.